\documentclass[twocolumn]{aastex62}

\newcommand{\teff}{$T_{eff}$}
\newcommand{\mjup}{$M_{\rm{Jup}}$~}

\newcommand{\hst}{\textit{HST~}}
\newcommand{\spitzer}{\textit{Spitzer~}}

\received{}
\revised{}
\accepted{\today}
\submitjournal{ApJ}

\shorttitle{WISEA~J0830+2837}
\shortauthors{Bardalez Gagliuffi et al.}

\begin{document}

\title{WISEA~J083011.95+283716.0: A Missing Link Planetary-Mass Object}

\correspondingauthor{Daniella C. Bardalez Gagliuffi}
\email{dbardalezgagliuffi@amnh.org}

\author[0000-0001-8170-7072]{Daniella C. Bardalez Gagliuffi}
\affil{American Museum of Natural History, 200 Central Park West, New York, NY 10024, USA}

\author[0000-0001-6251-0573]{Jacqueline K. Faherty}
\affil{American Museum of Natural History, 200 Central Park West, New York, NY 10024, USA}

\author[0000-0002-6294-5937]{Adam C. Schneider}
\affil{School of Earth and Space Exploration, Arizona State University, 781 Terrace Mall, Tempe, AZ 85287, USA}

\author[0000-0002-1125-7384]{Aaron Meisner}
\affil{NSF's National Optical-Infrared Astronomy Research Laboratory,  950 N Cherry Ave, Tucson, AZ 85719, USA}

\author[0000-0001-7896-5791]{Dan Caselden}
\affil{Gigamon Applied Threat Research, 619 Western Avenue, Suite 200, Seattle, WA 98104, USA}

\author[0000-0002-7630-1243]{Guillaume Colin}
\affil{Backyard Worlds: Planet 9}

\author[0000-0003-2236-2320]{Sam Goodman}
\affil{Backyard Worlds: Planet 9}

\author[0000-0003-4269-260X]{J.\ Davy Kirkpatrick}
\affil{IPAC, Mail Code 100-22, Caltech, 1200 E. California Blvd., Pasadena, CA 91125, USA}

\author[0000-0002-2387-5489]{Marc J. Kuchner}
\affil{NASA Goddard Space Flight Center, Exoplanets and Stellar Astrophysics Laboratory, Code 667, Greenbelt, MD 20771, USA}

\author[0000-0002-2592-9612]{Jonathan Gagn\'{e}}
\affil{Institut de Recherche sur les Exoplan\`{e}tes, Universit\'{e} de Montr\'{e}al, Pavillon Roger-Gaudry, PO Box 6128 Centre-Ville STN, Montreal QC  H3C 3J7, Canada}

\author[0000-0002-9632-9382]{Sarah E. Logsdon}
\affil{NSF's National Optical-Infrared Astronomy Research Laboratory,  950 N Cherry Ave, Tucson, AZ 85719, USA}

\author[0000-0002-6523-9536]{Adam J. Burgasser}
\affil{Center for Astrophysics and Space Sciences, University of California, San Diego, 9500 Gilman Drive, Mail Code 0424, La Jolla, CA 92093, USA}

\author[0000-0003-0580-7244]{Katelyn Allers}
\affil{Physics and Astronomy Department, Bucknell University, 701 Moore Ave, Lewisburg, PA 17837, USA}

\author[0000-0002-1783-8817]{John Debes}
\affil{Space Telescope Science Institute, Baltimore, MD 21218, USA}

\author[0000-0001-9209-1808]{John Wisniewski}
\affil{Homer L. Dodge Department of Physics and Astronomy, University of Oklahoma, 440 West Brooks Street, Norman, OK 73019, USA}

\author[0000-0003-4083-9962]{Austin Rothermich}
\affil{Physics Department, University Of Central Florida, 4000 Central Florida Boulevard, Building 12, 310, Orlando, FL 32816, USA}

\author[0000-0003-4714-3829]{Nikolaj S. Andersen}
\affil{Kolding Hospital, Department of Cardiology, Sygehus Lillebalt 24, 6000 Kolding, Denmark}

\author[0000-0001-5284-9231]{Melina Th\'evenot}
\affil{Backyard Worlds: Planet 9}

\author{Jim Walla}
\affil{Backyard Worlds: Planet 9}

\collaboration{Backyard Worlds: Planet 9 Collaboration}



\begin{abstract}
We present the discovery of WISEA~J083011.95+283716.0, the first Y dwarf candidate identified through the Backyard Worlds: Planet 9 citizen science project.  We identified this object as a red, fast-moving source with a faint $W2$ detection in multi-epoch \textit{AllWISE} and unWISE images. We have characterized this object with \spitzer and \textit{Hubble Space Telescope} follow-up imaging. With mid-infrared detections in \textit{Spitzer}'s \emph{ch1} and \emph{ch2} bands and flux upper limits in \hst $F105W$ and $F125W$ filters, we find that this object is both very faint and has extremely red colors ($ch1-ch2 = 3.25\pm0.23$\,mag, $F125W-ch2 \geq 9.36$\,mag), consistent with a T$_{eff}\sim300$\,K source, as estimated from the known Y dwarf population. A preliminary parallax provides a distance of $11.1^{+2.0}_{-1.5}$\,pc, leading to a slightly warmer temperature of $\sim350\,$K. The extreme faintness and red \hst and \spitzer colors of this object suggest it may be a link between the broader Y dwarf population and the coldest known brown dwarf WISE~J0855$-$0714, and highlight our limited knowledge of the true spread of Y dwarf colors. We also present four additional Backyard Worlds: Planet 9 late-T brown dwarf discoveries within 30\,pc.
\end{abstract}




\section{Introduction}\label{sec:intro}


With temperatures below $\sim500\,K$, Y dwarfs are the coldest products of star formation~\citep{2011ApJ...743...50C}, and are well within the temperature and mass range of old giant planets. 
Detecting Y dwarfs was a key goal of the \textit{Wide Field Infrared Survey Explorer}~(\textit{WISE};~\citealt{2010AJ....140.1868W}), with a mid-infrared filter designed specifically to cover the flux peak of these objects ($\lambda _{W2} = 4.6\,\mu$m). Several discoveries from this mission have reshaped the landscape of the solar neighborhood: WISE~J085510.83$-$071442.5 (hereafter: WISE~J0855$-$0714), the fourth closest system to the Sun 
($2.31\pm0.08$\,pc;~\citealt{2014ApJ...796....6L}) and the coldest known brown dwarf to date at $\sim$250\,K ~\citep{2014ApJ...786L..18L}; WISE~J104915. 57$?$531906.1, the closest brown dwarf binary to the Sun, also at 2\,pc~($2.02\pm0.019$\,pc;~\citealt{2014AandA...561L...4B}); and WISE J072003.20$-$084651.2AB, an M9+T5 dwarf spectral binary system at 6\,pc, which traversed the Oort cloud about 70,000 years ago at only 0.25\,pc~\citep{2014AandA...561A.113S,2015AJ....149..104B, 2015ApJ...800L..17M}, and whose T dwarf component is more massive than predicted from evolutionary models \citep{2019AJ....158..174D}. More recent discoveries have capitalized on multi-epoch \textit{WISE} photometry to reach fainter objects than previous detection thresholds allowed, like the $\sim270-360\,$K Y dwarf CWISEP~J193518.59$-$154620.3 (hereafter: CWISEP~J1935$-$1546;~\citealt{2019ApJ...881...17M}), and the $\sim310-360$\,K CWISEP~J144606.62$-$231717.8 (hereafter: CWISEP~J1446$-$2317)~\citep{2020ApJ...888L..19M, 2020ApJ...889...74M}. Nevertheless, the currently known Y dwarf population remains small, totaling only 28 objects~\citep{2011ApJ...743...50C,2011ApJ...730L...9L,2012ApJ...758...57L,2012ApJ...753..156K,2012ApJ...759...60T,2013ApJ...776..128K,2014ApJ...786L..18L,2014MNRAS.444.1931P,2014ApJ...796...39T,2015ApJ...803..102D,2015ApJ...804...92S,2018ApJ...867..109M,2018ApJS..236...28T,2019ApJ...881...17M,2020ApJ...888L..19M}. 

Even with this small sample,~\citet{2019ApJS..240...19K} placed initial constraints on the field luminosity and mass functions of brown dwarfs at cold temperatures. This study found that a single power law mass function extending down to 5\,\mjup 
provided a good fit to the number density of nearby brown dwarfs. However, sampling is particularly sparse at the lowest temperatures, with WISE~J0855$-$0714 being the only object in the $150-300$\,K temperature range. A larger volume-limited sample of Y dwarfs is needed to improve the statistical measurement of the luminosity function at these temperatures, in order to constrain the low-mass limit of star formation~\citep{2004ApJS..155..191B}. Similarly, a statistical sample of Y dwarfs is needed to characterize the composition, structure, and dynamics of low temperature atmospheres (e.g.~\citealt{2016ApJ...826L..17S,2014ApJ...793L..16F}), and possibly identify spectroscopic markers of different formation pathways between brown dwarfs and giant planets (e.g.~\citealt{2019ApJ...882L..29M}). 

Y dwarfs have complex atmospheres that resemble those of giant planets, with deep absorption features from CH$_4$, H$_2$O, and possibly  NH$_3$~(\citealt{2003ApJ...596..587B,2012ApJ...745...26B}), and ice water clouds~\citep{2014ApJ...787...78M}. Since they are not outshined by a host star, these objects generally are ideal proxies to conduct exoplanet atmospheric characterization studies. However, due to their intrinsic faintness, objects with spectral types later than Y1 either do not have spectroscopic measurements or poor signal-to-noise data that prevent the clear identification of molecular absorption lines and bands. For example, phosphine, salt, and sulfide clouds, similar to those found in the atmospheres of Jupiter~\citep{2006ApJ...648.1181V}, were predicted to affect the mid-infrared spectra of Y dwarfs~\citep{2012ApJ...756..172M}. However, phosphine was not identified in the M-band spectrum of the 250\,K WISE~J0855$-$0714~\citep{2018ApJ...858...97M,2016ApJ...826L..17S}, and neither were salt nor sulfide clouds, even though their presence is favored by atmospheric retrieval models~\citep{2019ApJ...877...24Z}. Y dwarfs are prime targets for medium resolution spectroscopy with the \textit{James Webb Space Telescope (JWST)} to solidly determine their atmospheric composition.


Despite the growth in the known population of cold brown dwarfs, there remain two brightness and temperature gaps between the majority of the known Y dwarfs, WISE~J0855$-$0714, and gas giants like Jupiter~($\sim125\,$K;~\citealt{1981JGR....86.8705H}). To look for the faintest objects missed in previous searches, the Backyard Worlds: Planet 9 citizen science project (BYW:P9) is harnessing the power of visual inspection of multi-epoch and multi-wavelength imaging. The goal of this project is to identify the faintest, coldest, and fastest-moving sources in the Solar neighborhood through time-resolved WISE coadded animations (unWISE;~\citealt{2017AJ....154..161M,2017AJ....153...38M,2018AJ....156...69M}). BYW:P9 was launched in February 2017~\citep{2017ApJ...841L..19K} as part of the Zooniverse ecosystem. Since then, over 150,000 users around the world have participated in the visual identification and classification of new brown dwarf candidates through the Backyard Worlds interface\footnote{\url{www.BackyardWorlds.org}}. Prior discoveries include the lowest binding energy ultracool binary in the field~\citep{2019arXiv191104600F}, a white dwarf with infrared excess~\citep{2019ApJ...872L..25D}, and two T-type subdwarfs~(Schneider et al., submitted).


In this paper, we present the discovery and space-based follow-up observations of 5 new low-temperature brown dwarfs identified by Backyard Worlds: Planet 9 citizen scientists, including one Y dwarf. Section~\ref{sec:sample} describes the five targets of our sample. Section~\ref{sec:obs} describes the observations with the \textit{Hubble} and \textit{Spitzer} space telescopes and our photometric measurements. Section~\ref{sec:analysis} shows our parallax calculation and estimated quantities based on our infrared color analysis. Section~\ref{sec:discussion} discusses the implications of our discoveries for our understanding of the T/Y transition, the census of the Solar neighborhood, and brown dwarf formation at the lowest masses.


\section{Target Sample}\label{sec:sample}

The \textit{WISE} observing strategy consists of scanning the entire sky every six months, with over 12 exposures per visit. Co-adding exposures from the full cryogenic \textit{WISE}~\citep{2010AJ....140.1868W}, NEOWISE~\citep{2011ApJ...743..156M}, and NEOWISE Reactivation missions~(NEOWISER;~\citealt{2014ApJ...792...30M}), led to the generation of full-depth, ``unWISE'' images in $W1$ and $W2$ filters ~\citep{2017AJ....153...38M,2018AJ....156...69M}. These images reach $\sim1.3$\,mag fainter than single exposures, and are used for the identification of faint, fast-moving sources through BYW:P9. 

Our sample contains five sources identified by citizen scientists from the BYW:P9 collaboration due to their red color in $W1-W2$ and fast proper motion in the 5.5-year baseline of unWISE images~\citep{2018AJ....156...69M,2018RNAAS...2..202M}. None of these objects are detected in the near-infrared Two Micron All Sky Survey (2MASS;~\citealt{2006AJ....131.1163S}), the UKIRT Hemisphere Survey (UHS;~\citealt{2018MNRAS.473.5113D}), or the VISTA Hemisphere Survey (VHS;~\citealt{2013Msngr.154...35M}). In the mid-infrared, our five sources have only flux upper limit AllWISE photometry in $W1$ and detections with large uncertainties in $W2$, leading to uncertain $W1-W2$ colors that nevertheless indicate extremely faint and red objects.

The precision and accuracy of the photometry and proper motions of these objects are significantly improved in the  CatWISE Preliminary catalog~\citep{2019arXiv190808902E} relative to AllWISE. CatWISE is a new catalog resulting from running the AllWISE software on unWISE images using NEOWISE epochs, yielding 10X better per-coordinate proper motion uncertainty at $W1\sim1$5\,mag, and is 3 magnitudes more sensitive than AllWISE at 100\,$mas/yr$. Since AllWISE proper motions are generally unreliable at $W2\geq13.5\,$mag~\citep{2016ApJS..224...36K}, and have been superseded by CatWISE, we present only CatWISE photometry and proper motions in  Table~\ref{tab:wise} and use these measurements in the analysis that follows.

\begin{figure}
\figurenum{1}
\centering
\includegraphics[width=0.48\textwidth]{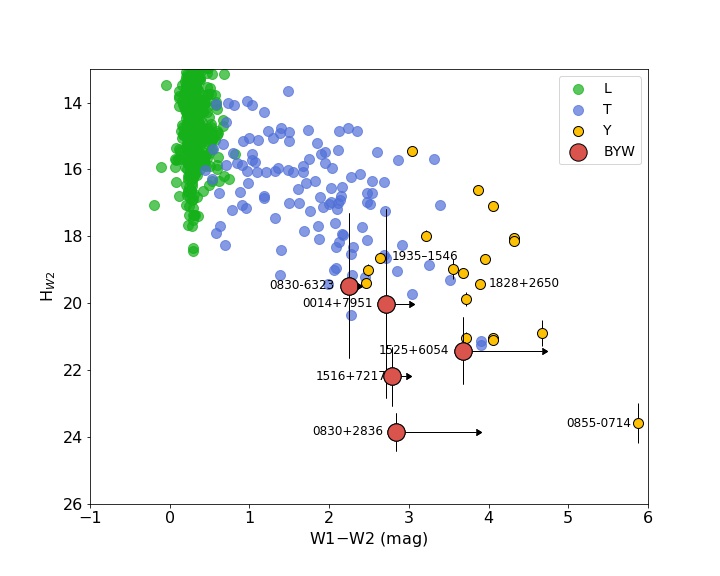}
\caption{Reduced proper motion diagram in $W2$ band from AllWISE. All five objects in our sample are flux upper limits in $W1$ from AllWISE. Background M, L, and T dwarfs come from~\citet{2009AJ....137....1F,2012ApJ...752...56F}. Y dwarfs are compiled from~\citet{2019ApJS..240...19K}.}\label{fig:rpm}
\end{figure}

Using multi-epoch unWISE data~\citep{2018AJ....156...69M}, we measured proper motions for each source and placed them in a reduced proper motion diagram, traditionally used to distinguish faint main sequence stars from subdwarfs and cool, white dwarfs if parallaxes are absent, acting as a proxy for absolute magnitude~\citep{1922LicOB..10..135L}. For our BYW:P9 \textit{HST} follow-up, we prioritized five sources that overlap with the known Y dwarf population (see Figure~\ref{fig:rpm}). These sources were also observed as part of our \textit{Spitzer} follow-up program. In the next sections, we characterize these sources with both sets of photometry.

\begin{deluxetable*}{lccccccc}
\tabletypesize{\scriptsize}
\tablewidth{0pt}
\tablenum{1}
\tablecolumns{7}
\tablecaption{CatWISE photometry and proper motions.\label{tab:wise}}
\tablehead{ 
\colhead{Source} & 
\colhead{CatWISE Designation} & 
\colhead{$W1$ (mag)} & 
\colhead{$W2$ (mag)} & 
\colhead{$W1-W2$ (mag)} & 
\colhead{$\mu_{\alpha}$ (mas/yr)} & 
\colhead{$\mu_{\delta}$ (mas/yr)} & 
\colhead{$\mu_{total}$ (mas/yr)}}
\startdata
WISEA~J001449.96+795116.1 &  CWISEP~J001450.72+795116.0 &
18.72$\pm$0.28 &  16.00$\pm$0.06 &     2.72$\pm$0.28 &   531.2$\pm$89.2 &   -20.7$\pm$77.8 &  531.6$\pm$89.2 \\ 
WISEA~J083011.95+283716.0\tablenotemark{*} & CWISEP~J083011.94+283716.2 &  $\geq$18.0 &  16.639$\pm$0.188 &     $\geq$1.33 &    -730.40$\pm$290.7 &  -1161.20$\pm$367.0 &  1371.43$\pm$468.18 \\ 
  & CWISEP~J083011.94+283706.1 &  $\geq$18.9 &  16.052$\pm$0.092 &     $\geq$2.9 &    54.3$\pm$153.5 &  -314.3$\pm$154.1 &  319.0$\pm$154.1 \\ 
WISEA~J083019.97$-$632305.4 & CWISEP~J083019.95-632304.2 &  18.12$\pm$0.10 &  15.87$\pm$0.04 &  2.25$\pm$0.10 &   -81.7$\pm$65.4 &   375.0$\pm$64.7 &  383.8$\pm$64.7 \\ 
WISEA~J151620.39+721745.4 & CWISEP~J151620.00+721747.9 &   18.91$\pm$0.17 &  16.13$\pm$0.05 & 2.79$\pm$0.18 &  -562.2$\pm$89.9 &   560.8$\pm$78.9 &  794.1$\pm$84.5 \\ 
WISEA~J152529.09+605356.5  & CWISEP~J152528.90+605359.0 &  $\geq$19.7 &  15.99$\pm$ 0.05 & $\geq$3.7 &  -394.2$\pm$85.1 &   763.7$\pm$85.6 &  859.4$\pm$85.5 \\
\enddata
\tablenotetext{*}{Two CatWISE entries are found when searching for the AllWISE source WISEA~J083011.95+283716.0. This object has such a high proper motion that the CatWISE pipeline confused it as two sources. The photometry and proper motion reported in CatWISE for this object are unreliable. Our \spitzer photometry (Table~\ref{tab:phot}) and astrometry (Table~\ref{tab:astrometry_0830}) supersede the values reported in this Table for this object.}
\tablecomments{Object names are from AllWISE, magnitudes and proper motions are updated from CatWISE. For CatWISE, we report photometry from the \texttt{w1mpro\_pm} and \texttt{w2mpro\_pm} keywords and their associated uncertainties.}
\end{deluxetable*}

\section{Observations}\label{sec:obs}

\subsection{WFC3/HST Photometry}

Our targets were observed between UT 11 August 2018 and UT 16 February 2019 in 5 \textit{HST} orbits under GO program 15468 (PI: Faherty). All five targets were observed with both F105W ($\lambda_c = 1055.2$\,nm) and F125W ($\lambda_c = 1248.6$\,nm) filters on the Wide Field Camera 3 (WFC3;~\citealt{2008SPIE.7010E..1FM}), which roughly coincide with the $Y$ and $J$ near-infrared bands. All of the observations were done in MULTIACUUM mode with 4 dither positions per source per filter in a $123''\times135''$ field of view. Total exposure times are listed in Table~\ref{tab:obs}. Scheduled observations of WISEA~J1516+7217 on UT 2018 November 5 were taken on gyros only, after failing to acquire a guide star, but these were rescheduled and successfully acquired on UT 2019 February 16. The individual exposures per filter were combined, cleaned of cosmic rays, and registered to the World Coordinate System (WCS) with the \texttt{tweakreg} and \texttt{astrodrizzle} routines available in the AstroDrizzle software~\citep{2012drzp.book.....G}.

\begin{deluxetable*}{lccccc}
\tabletypesize{\scriptsize}
\tablewidth{0pt}
\tablenum{2}
\tablecolumns{6}
\tablecaption{Summary of \hst and \spitzer observations.\label{tab:obs}}
\tablehead{
\colhead{Source} & 
\colhead{R.A.} & 
\colhead{Dec.} & 
\colhead{Obs. Date} &
\colhead{Total Exp. Time (s)} & 
\colhead{Filter}}
\startdata
\cutinhead{\hst}
WISEA~J0014+7951 & 00 15 01.32 & +79 51 08.08 & 2018  Aug 11 & 1412 & F105W\\
 & & & & 1412 & F125W\\
WISEA~J0830+2837 & 08 30 11.89 & +28 36 58.23 & 2018 Sep 29 & 1212 & F105W\\
 & & & & 1312 & F125W\\
WISEA~J0830$-$6323 & 08 30 19.95 & -63 22 59.97 & 2018 Aug 29 & 1312 & F105W\\
& & & & 1412 & F125W\\
WISEA~J1516+7217 & 15 16 19.43 & +72 17 53.81 & 2019 Feb 16 & 1412 & F105W\\
& & & & 1412 & F125W\\
WISEA~J1525+6053 & 15 25 28.65 & +60 54 03.32 & 2018 Nov 5 & 1312 & F105W\\
& & & & 1412 & F125W\\
\cutinhead{\spitzer}
WISEA~J0014+7951 & 00 14 49.96 & +79 51 16.2 & 2018 Dec 16 & 378 & ch1\\
 & & & & 378 & ch2\\
WISEA~J0830+2837 & 08 30 11.96 & +28 37 16.0  & 2019 Feb 21 & 378 & ch1\\
 & & & & 378 & ch2\\
WISEA~J0830$-$6323 &  08 30 19.98 & -63 23 05.5  & 2018 Aug 14 & 378 & ch1\\
& & & & 378 & ch2\\
WISEA~J1516+7217 & 15 16 20.40 & +72 17 45.5  & 2019 Feb 11 & 378  & ch1\\
& & & & 378 & ch2\\
WISEA~J1525+6053 & 15 25 29.10 & +60 53 56.6  & 2018 Oct 21 & 378 & ch1\\
& & & & 378 & ch2 \\
\enddata
\end{deluxetable*}

Photometry was calculated on the drizzled images using a custom function\footnote{Found at \url{https://github.com/daniellabardalezgagliuffi/HSTphotometry}} built with routines from the \texttt{photutils} Python package. Centroids were determined using the \texttt{DAOStarFinder} Python routine based on the DAOFIND algorithm~\citep{1987PASP...99..191S}. This algorithm uses a 2D Gaussian kernel to search for local maxima with a peak amplitude greater than a given threshold. We chose a full-width half maximum of 3 pixels and a threshold of 5 times the median absolute deviation in a 50-by-50 pixel cut-out of the science images centered on the targets. This method yielded more accurate results than other 1-D and 2-D Gaussian fitting techniques (e.g., \texttt{photutils.centroids}).

WFC3 pixels become correlated after the drizzling process, so rather than calculating the background flux with an annulus around the target, we followed the prescription of~\citet{2015ApJ...804...92S}. The contribution of the sky background was estimated by randomly placing 10,000 apertures on the image, and calculating aperture photometry on them. The radius for these apertures was $0\farcs4$ in flux units of $e^{-}/s$, thus matching the one used for aperture photometry on the sources, as defined in the \textit{WFC3} Data Handbook~\citep{WFC3DataHandbook}. The array of background count values was sigma-clipped to $3\sigma$ to avoid contribution from apertures containing stars or anomalous negative-count pixels. The median and standard deviation of the background are used as the value and uncertainty in the sky contribution.

Aperture photometry on the stellar sources was calculated using the same radius of $0\farcs4$. The final source flux was obtained by subtracting the background flux from the aperture flux, since both aperture areas were the same. Magnitudes in the Vega system were directly calculated from the fluxes using zero points of 25.4523 mag and 25.1439 mag for the F105W and F125W filters, respectively, as defined by the \textit{HST}/WFC3 data handbook~\citep{WFC3DataHandbook}. Uncertainties on the stellar flux were estimated following the DAOPHOT photometry error method~\citep{1987PASP...99..191S} and includes Poisson error contributions from stellar and background counts and read error from the detector:

\begin{equation}
\sigma_{tot} = \frac{\sqrt{\frac{F\Delta t_{exp}}{g_eff} + A (\sigma_{bkg}\Delta t_{exp})^2 + A^2 \frac{(\sigma_{bkg} \Delta t_{exp})}{N_{sky}}}}{{\Delta t_{exp}}}\\
\end{equation}

where $F$ is the raw aperture flux (including both stellar and background contributions), $\Delta t_{exp}$ is the exposure time, $g_eff$ is the effective gain, $A$ is the aperture area, $\sigma_{bkg}$ is the standard deviation of the background in $e^{-}/s$, and $N_{sky}$ is the number of sky pixels used to estimate the background. Since the WFC3 images come in units of $e^{-}/s$, the effective gain can be estimated as the total exposure time~\footnote{See~\url{https://photutils.readthedocs.io/en/stable/api/photutils.utils.calc_total_error.html} for more details.}, as extracted from the FITS headers. For a total uncertainty per pixel per second, this equation can be reduced to:

\begin{equation}
\sigma_{tot} = \sqrt{\frac{F}{\Delta t_{exp}} + 2\sigma_{bkg}^2}
\end{equation}  



For the case of WISEA~J0830+2837, which is undetected in both {\it HST} filters, we estimate a $3\sigma$ upper limit on flux from the standard deviation of the flux from the 10,000 background apertures.
 
The $HST$ images of WISEA~J0014+7951 reveal a nearby fainter source $1\farcs2$ away at a position angle of $231.52^{\circ}$ East of North. We measured $F105W$ and $F125W$ magnitudes of 24.07$\pm$0.265\,mag and 23.59$\pm$0.271\,mag, respectively, leading to an \hst color of 0.48$\pm$0.38\,mag. This source is $\sim$3\,mag fainter than our WISEA~J0014+7951 target in both bands ($F105W = 20.924\pm0.015$\,mag, $F125W = 19.993\pm0.008$\,mag), and its \hst color is roughly half a magnitude bluer than the \hst color of WISEA~J0014+7951 ($F105W-F125W = 0.931\pm0.017$\,mag, see Table~\ref{tab:phot}). We hypothesize, that if this object were a true companion to the T8 WISEA~J0014+7951, by definition it would be cooler, later-type, and hence redder (c.f.~\citealt{2015ApJ...804...92S, 2016ApJ...823L..35S, 2019ApJS..240...19K}). However, given the bluer \hst color, it is unlikely that these two objects are associated. Unfortunately, these two sources are blended in our \spitzer images, so we are unable to measure a second color to better characterize this object. 

\subsection{Spitzer photometry}


In addition to the \hst photometry, we also obtained \spitzer photometry for these five targets, as part of program ID 14076 (PI: Faherty).  The observations followed a 16-point spiral dither pattern with 30\,s exposures per frame. Data were acquired with both $ch1$ and $ch2$ filters. Readout was done in full array mode. 

Aperture photometry was measured using the \spitzer MOsaicker and Point Source EXtractor with point-source extraction package (MOPEX/APEX;~\citealt{2005PASP..117.1113M}), and can be found in Table~\ref{tab:obs}. Specifically, we used the corrected basic calibrated data (CBCD) frames to build custom mosaics from which the fluxes were measured using a 4-pixel aperture.  These raw fluxes were converted to magnitudes by applying an aperture correction and comparing to the published $ch1$ and $ch2$ flux zero points, as described in section 5.1 of \citet{2019ApJS..240...19K}.


\begin{deluxetable*}{lccccc}
\tabletypesize{\scriptsize}
\tablewidth{0pt}
\tablenum{3}
\tablecolumns{8}
\tablecaption{\hst and \spitzer photometry.\label{tab:phot}}
\tablehead{
\colhead{Parameter} &
\colhead{WISEA~J0014+7951} & 
\colhead{WISEA~J0830+2837} & 
\colhead{WISEA~J0830$-$6323} & 
\colhead{WISEA~J1516+7217} & 
\colhead{WISEA~J1525+6053}} 
\startdata
\hst Ap. $F105W$ (mag) & 20.924$\pm$0.015 & $\geq$25.4 & 20.225$\pm$0.010 & 21.470$\pm$0.013  & 21.173$\pm$0.013 \\
\hst Ap. $F125W$ (mag) & 19.993$\pm$0.008 & $\geq$25.2 & 19.297$\pm$0.004 & 20.557$\pm$0.007 & 20.233$\pm$0.006\\
\hst Ap. $F105W-F125W$ (mag) & 0.931$\pm$0.017  & $-$ & 0.928$\pm$0.011 &  0.913$\pm$0.015  & 0.940$\pm$0.014\\
\spitzer PRF $ch1$  (mag)& 17.727$\pm$0.068 & 19.089$\pm$0.232 & 17.514$\pm$0.058 & 18.121$\pm$0.096 & 18.000$\pm$0.084\\
\spitzer PRF $ch2$  (mag)& 15.880$\pm$0.021 &15.837$\pm$0.021  & 15.682$\pm$0.019 & 15.946$\pm$0.021 & 15.874$\pm$0.021\\
\spitzer PRF $ch1-ch2$  (mag)&  1.847$\pm$0.071 & 3.252$\pm$0.233 & 1.832$\pm$0.061 & 2.175$\pm$0.098 & 2.126$\pm$0.087\\
\spitzer Ap. $ch1$   (mag)& 17.777$\pm$0.050 & 19.110$\pm$0.166 & 17.479$\pm$0.040 & 18.220$\pm$0.072 & 18.048$\pm$0.063\\
\spitzer Ap. $ch2$  (mag)& 15.845$\pm$0.017 & 15.854$\pm$0.018 & 15.669$\pm$0.017 & 15.947$\pm$0.018 & 15.893$\pm$0.018\\
\spitzer Ap. $ch1-ch2$  (mag)& 1.932$\pm$0.053 & 3.256$\pm$0.167 & 1.810$\pm$0.043 & 2.273$\pm$0.074 & 2.155$\pm$0.655\\
Discoverer\tablenotemark{*} & 1, 2, 3, 4, 5, 7 & 1, 2 & 2, 5 & 2 & 2, 6\\
 \enddata
\tablenotetext{*}{(1) Dan Caselden; (2) Guillaume Colin; (3) Sam Goodman;  (4) Austin Rothermich; (5) Nikolaj Stevnbak; (6) Melina Thevenot; (7) Jim Walla.}
\end{deluxetable*}

\section{Analysis}\label{sec:analysis}

\begin{figure*}
\figurenum{2}
\centering
\includegraphics[trim=80 0 0 10 ,width=1.1\textwidth]{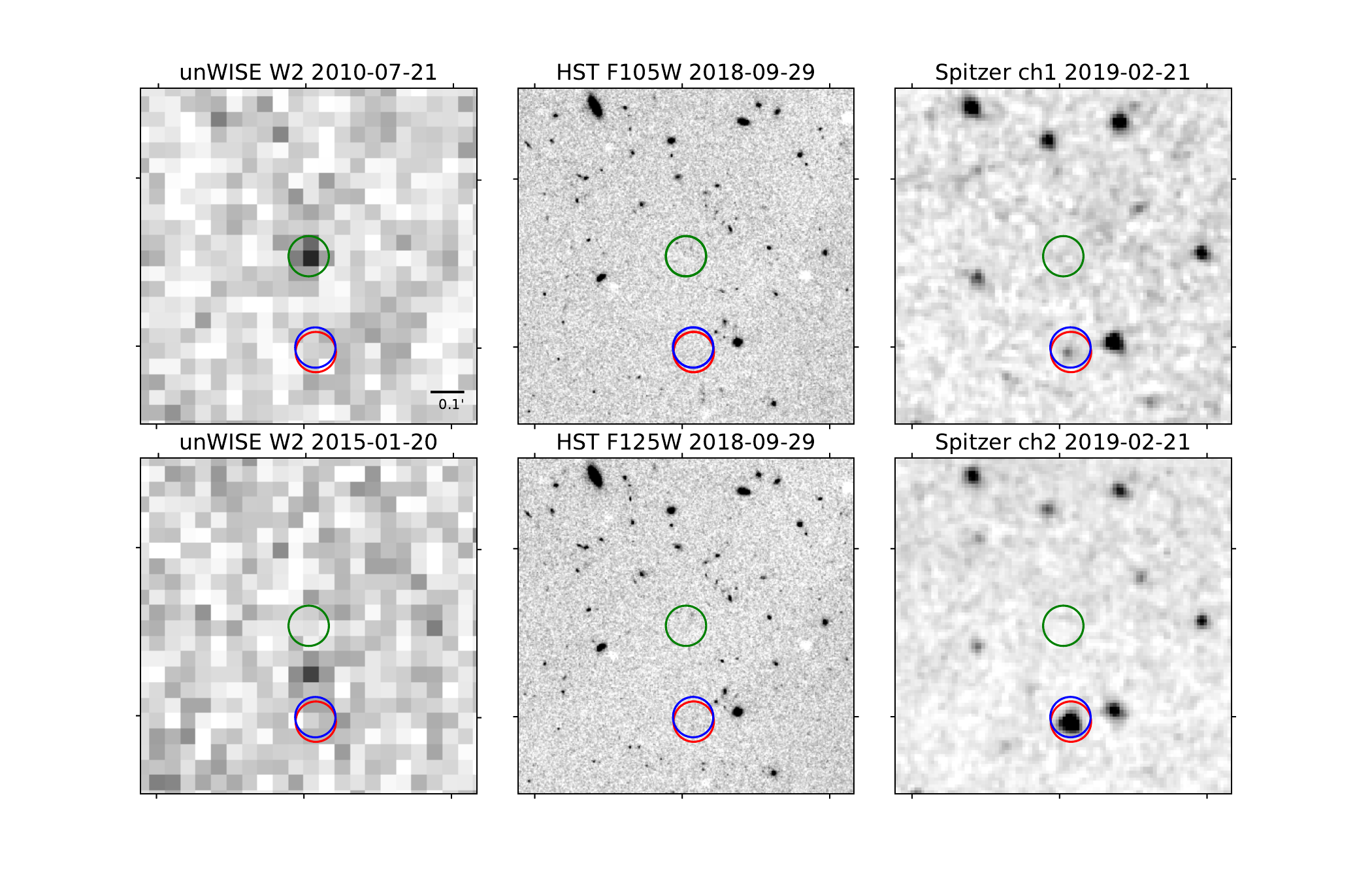}
\caption{Finder charts for WISEA~J0830+2837 of 1 arcminute per side. All frames are centered on the source at the first unWISE epoch (top left) and aligned North up, East to the left. unWISE positions are shown in green, \hst positions in blue, and \spitzer positions in red. Significant proper motion can be seen between the two epochs of unWISE images. For \hst images, we estimated the location of the source based on the \spitzer location and proper motion. The source is significantly brighter in \spitzer ch2 compared to ch1, and not detected in the NIR \hst filters, signaling an extremely cold object.  \label{fig:poststamps}}
\end{figure*}

\subsection{Preliminary parallax of WISEA~J0830+2837}

\spitzer $ch2$ imaging of WISEA~J0830+2837 was obtained in 2019 Feb via program 14076 (PI: Faherty) and in 2019 Jul-Aug via program 14224 (PI: Kirkpatrick). Astrometry was measured from the \spitzer images using the methodology described in section 5.2 of \citet{2019ApJS..240...19K} with a couple of exceptions. First, we used re-registration stars sampling down to smaller S/N values (S/N=30) in order to have a larger selection of objects per frame that match {\it Gaia} DR2 sources with full astrometric solutions. Second, we used the full astrometric solutions of these re-registration stars to place each frame on an absolute astrometric grid -- that is, we move each {\it Gaia} source to its expected position at that epoch and re-register the frame to that epoch-specific reference. This enables us to measure an absolute parallax, obviating the need to apply an ad hoc relative-to-absolute adjustment later.

After the original photometric follow-up in program 14076, we had only one additional visibility window available to us in program 14224 with which to measure astrometry for this object, because the \spitzer mission ceased operations on 30 Jan 2020. The data from programs 14076 and 14224 sample opposite sides of the parallactic ellipse, since they are separated by $\sim$6 months, but at least one other epoch of data is needed to disentangle proper motion from parallax. We therefore used source detections from the unWISE epochal coadds~\citep{2014AJ....147..108L} in a region around WISE~0830+2837, re-registered their positions to the {\it Gaia} DR2 reference frame, and used the resulting re-registered unWISE astrometry to provide eleven additional epochs spanning Apr 2010 to Oct 2018. (See~\citealt{2020ApJ...889...74M} for more details on the process.) These unWISE data were then associated with the appropriate {\it XYZ} position of the Earth at the mean time of each of the eleven unWISE epochs in the ICRS reference frame. A combined proper motion + parallax solution was fit using the methodology outlined in Section 5.2.3 of \citet{2019ApJS..240...19K}.

\begin{figure*}
\figurenum{3}
\centering
\includegraphics[width=0.75\textwidth]{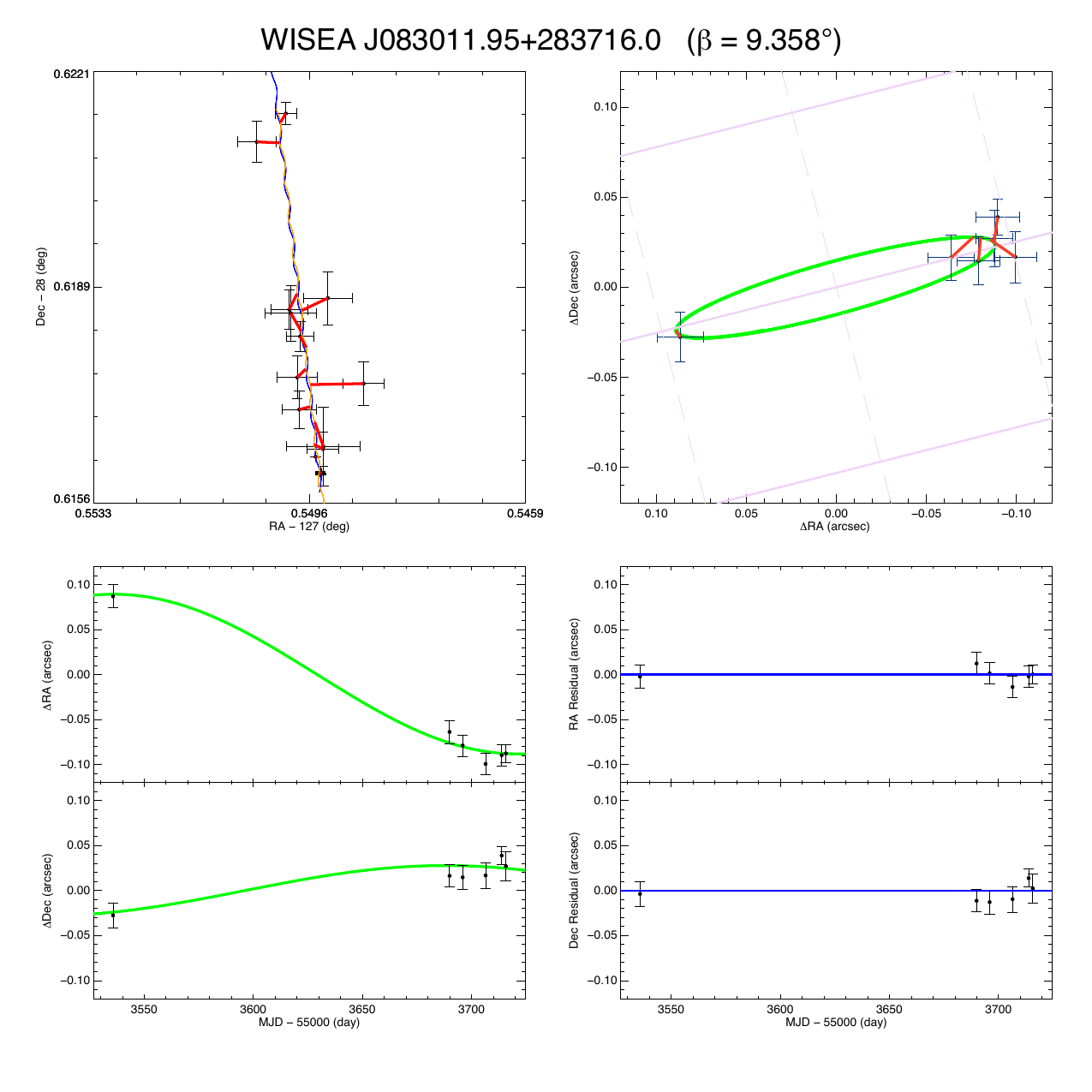}
\caption{Our fit to the astrometry of WISEA~0830+2837. The full solution and full data sets are show in the upper left panel, where the unWISE data points are the black points with large error bars and the \spitzer data points are the black points with the much smaller error bars. The blue curve is the expected astrometric path of the object as seen from \spitzer, and the orange curve is the path as seen from the Earth. Each of the observed data points is connected by a red line back to the spot on the relevant curve that has the same time stamp. The other panels -- the parallax-only solution (green curve, upper right), residuals around the parallax solution (lower left), and residuals around the full solution (lower right) -- show only the \spitzer data points, for clarity. \label{fig:astrometry_0830}}
\end{figure*}

Our resulting solution is given in Table~\ref{tab:astrometry_0830} and illustrated in Figure~\ref{fig:astrometry_0830}. The parallactic solution should be considered preliminary and somewhat fragile because there is a single high-quality data point anchoring one side of the parallactic ellipse. This is further demonstrated by the large parallactic error of $\sim$15\%.

\begin{deluxetable}{lc}
\tabletypesize{\footnotesize}
\tablenum{4}
\tablecaption{Preliminary \textit{WISE} and \spitzer Parallax and Motion Fit for WISE~0830+2837\label{tab:astrometry_0830}}
\tablehead{\colhead{Parameter} & \colhead{Value}}
\startdata
RA at t$_0$ & 127.549578(114.9)\\
Dec at t$_0$ & 28.617508(74.3)\\
t$_0$ (MJD) & 57762.10\\
$\pi_{abs}$ (mas) & 90.6$\pm$13.7\\
$\mu_{RA}$ (mas yr$^{-1}$) & -233.3$\pm$48.6\\
$\mu_{Dec}$ (mas yr$^{-1}$) & -2040.8$\pm$29.9\\
$\chi^2$ & 20.370\\                   
Dof & 29\\
Red. $\chi^2$ & 0.702\\   
\enddata
\end{deluxetable}

We find that the parallax is $90.6\pm13.7\,mas$, implying a distance of $11.1^{+2.0}_{-1.5}$\,pc. This parallax indicates an absolute magnitude in $ch2$ of $15.61\pm0.05$\,mag, which leads to a temperature range of $281-427$\,K following the~\citet{2019ApJS..240...19K} empirical relations.


\subsection{Near and Mid-Infrared Colors and Estimated Quantities}

\begin{figure}
\figurenum{4}
\includegraphics[width=0.48\textwidth]{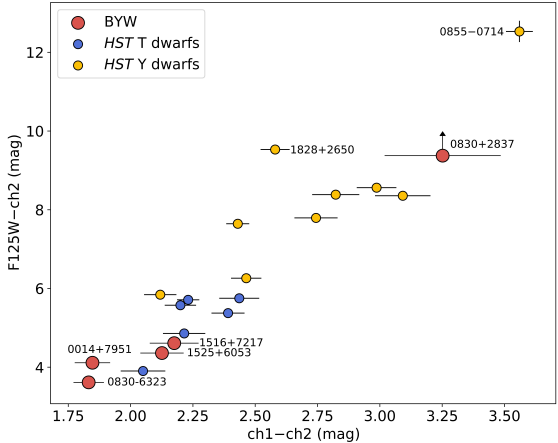}
\caption{Color-color diagram of \spitzer ch1$-$ch2 color against F125W$-ch2$. Our Backyard Worlds sample is colored in red. \hst and \spitzer photometry for T and Y dwarfs from~\citet{2015ApJ...804...92S, 2016ApJ...823L..35S} are shown as blue and yellow circles, respectively. The \spitzer photometry of WISEA~J0830+2837 places it solidly in the Y-dwarf populated region, with extremely red colors bridging the gap between WISE~J0855$-$0714 and the rest of the Y dwarf population.\label{fig:colorcolor}}
\end{figure}

\begin{figure}
\figurenum{5}
\includegraphics[width=0.48\textwidth]{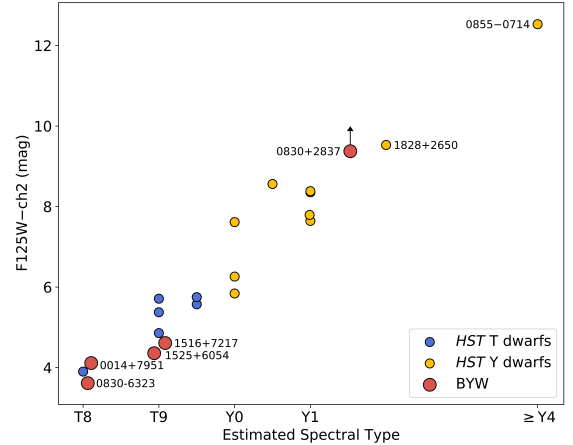}
\caption{Estimated spectral type vs. \hst and \textit{Spitzer} color. Our Backyard Worlds/\spitzer sample is colored in red. T and Y dwarfs from~\citep{2016ApJ...823L..35S} are shown in blue and yellow, respectively. Error bars for our sample are smaller than the symbol size, on average $\sim0.05$\,mag. \label{fig:F125ch2spt}}
\end{figure}

Equipped with \textit{HST} and \textit{Spitzer} photometry, we can characterize these objects in the context of other T and Y dwarfs. We have gathered colors from 14 late-T and Y dwarfs with \hst and \spitzer photometry in the literature \citep{2015ApJ...804...92S, 2016ApJ...823L..35S} and used these as a comparison sample. Figure~\ref{fig:colorcolor} shows our five targets in $ch1-ch2$ vs $F125W-ch2$ color-color space. Based on the colors of the comparison sample, it appears that all the objects in our sample, except for WISEA~J0830+2837, are late-T dwarfs. WISEA~J0830+2837 appears to be an early-Y dwarf (Table~\ref{tab:phot}, and Figure~\ref{fig:F125ch2spt}). 



Following the~\citet{2019ApJS..240...19K} relations for absolute magnitude in $ch2$, we estimate distances of $23-28$\,pc for our T dwarf candidates, and $8.7^{+2.3}_{-1.9}$\,pc for WISEA~J0830+2837, with Monte Carlo uncertainties at the $16^{th}$ and $84^{th}$ percentiles. This photometric distance is consistent with the parallax measurement within uncertainties. However, it is worth noting that the empirical relations rely on one data point beyond $ch1-ch2\sim3.5\,$mag or $M_{ch2}\sim16\,$mag, corresponding to WISE~J0855$-0714$, hence making any color-based estimation extremely tentative.

From every standpoint, WISEA~J0830+2837 is an outlier in our sample. This object is undetected in both \hst $F105W$ and $F125W$ filters, thus we only have flux upper limits. In \spitzer colors however, this source is detected with moderately large error bars. It is the faintest object in $ch1$ in our sample, and the second brightest in $ch2$. The \spitzer $ch1-ch2$ colors for our five targets are in the $1.8-2.1$\,mag range with the exception of WISEA~J0830+2837, which has the reddest $ch1-ch2$ color of the sample ($ch1-ch2 = 3.25\pm0.23$\,mag). Its color is comparable to CWISEP~J1935$-$1546~($ch1-ch2 = 2.984\pm0.034$\,mag,~\citealt{2020ApJ...889...74M}), and within $1\sigma$ of the \spitzer color of WISE~J0855$-$0714~($\Delta~ch1-ch2 = 0.3\pm0.2$\,mag). Additionally, from the \spitzer colors of WISEA~J0830+2837, we can estimate an effective temperature of $\sim300\,$K for this object~\citep{2019ApJS..240...19K}, also analogous to CWISEP~J1935$-$1546 with a temperature ($270-360\,$K). These two objects appear to be ``missing links'' filling the gap between most known Y dwarfs and WISE~J0855$-$0714.


Based on parallax and proper motion of WISEA~J0830+2837 with \spitzer, we can estimate a tangential velocity of $V_{tan} = 107.9\pm16.67$\,km/s, which is significantly higher than the median $V_{tan}$ and velocity dispersion ($\sigma_{tan}$) for the nearby T dwarf population (31\,km/s and 20\,km/s, respectively;~\citealt{2012ApJ...752...56F}). The $V_{tan}$ of this object is also higher than that of comparable objects, such as the Y dwarf WISE J163940.83?684738.6~\citep{2012ApJ...759...60T}, with a $V_{tan} = 73\pm8$\,km/s. The high $V_{tan}$ of WISEA~J0830+2837 suggests a kinematically old age. Assuming an age between 1-10\,Gyr for this object, we estimate a mass of 4-13\,\mjup~using the~\citet{2015AandA...577A..42B} evolutionary models. Therefore, WISEA~J0830+2837 is likely a planetary-mass object. 

\begin{deluxetable}{lcccc}
\tabletypesize{\scriptsize}
\tablewidth{0pt}
\tablenum{5}
\tablecolumns{5}
\tablecaption{Estimated physical properties of our sample based on \spitzer colors and parallax measurement for WISEA~J0830+2837.\label{tab:props}}
\tablehead{
\colhead{Source} & 
\colhead{Photometric Type} &
\colhead{$T_{\rm{eff}}$ (K)} & 
\colhead{Distance (pc)}} 
\startdata
WISEA~J0014+7951 & T8  & 659$\pm$85  & 28$\pm$4\\ 
WISEA~J0830+2837 & $\geq$Y1 &  303$\pm$87 &  9$\pm$2\\
 & $\geq$Y1\tablenotemark{*} &  354$\pm$73\tablenotemark{*} &  11.1$^{+2.0}_{-1.5}$\tablenotemark{*}\\
WISEA~J0830$-$6323  & T8 &  664$\pm$84 &  26$\pm$4 \\
WISEA~J1516+7217  & T9  & 548$\pm$87 &  24$\pm$4 \\ 
WISEA~J1525+6053  & T9 &   563$\pm$86 &  24$\pm$4 \\
 \enddata
 \tablenotetext{*}{Calculated distance, and estimated photometric type and effective temperature from \spitzer parallax measurement. }
\end{deluxetable}


\section{Discussion}\label{sec:discussion}

\begin{figure}
\figurenum{6}
\includegraphics[width=0.48\textwidth]{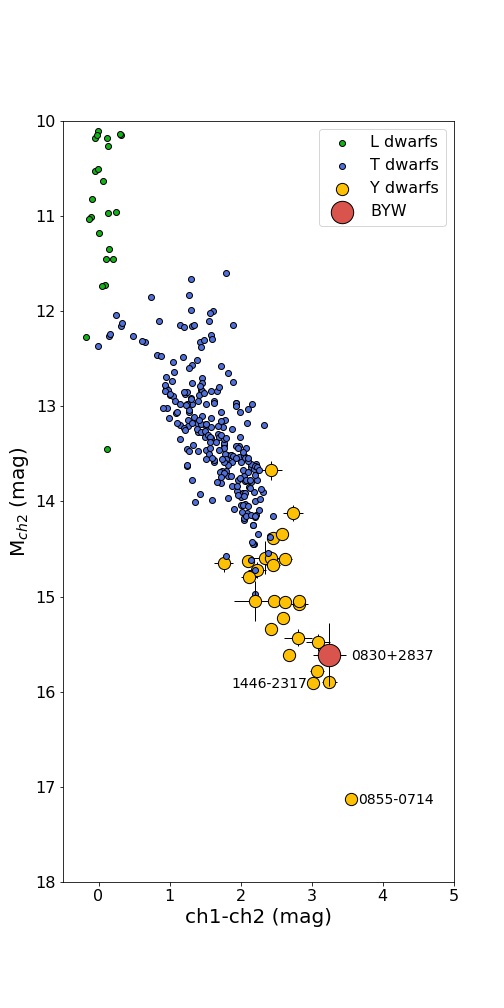}
\caption{Color-magnitude diagram in \spitzer IRAC $ch2$-band. L dwarf parallaxes and $ch2$ photometry come from the Database of Ultracool Parallaxes~\citet{2012ApJS..201...19D, 2013Sci...341.1492D, 2016ApJ...833...96L}; T and Y dwarf parallaxes come from~\citet{2019ApJS..240...19K}. WISEA~J0830+2837 joins CWISEP~J1935$-$1546 and CWISEP~J1446$-$2317 on a small, yet growing sample of ``missing link'' objects bridging the known Y dwarf population to the coldest brown dwarf known, WISE~J0855$-$0714.\label{fig:CMD}}
\end{figure}

\subsection{The low mass end of the substellar IMF}

As shown in Figure~\ref{fig:CMD}, WISEA~J0830+2837 is one of the faintest and reddest objects in the Y dwarf population. Similar in brightness and \spitzer color are CWISEP~J1935$-$1546~\citep{2019ApJ...881...17M} and CWISEP~J1446$-$2317~\citep{2020ApJ...888L..19M, 2020ApJ...889...74M}. These three objects constitute a small sample bridging the known Y dwarf population to WISE~J0855$-$0714~\citep{2014ApJ...786L..18L}.  Constraining the initial mass function (IMF) observationally is crucial to understanding the low-mass limit of brown dwarf formation, and Y dwarfs are the lowest-mass piece in the present-day luminosity function that maps onto an IMF. Perhaps the most fundamental open question in brown dwarf science is how these low-mass objects form. Brown dwarfs most likely form in a process similar to an extension of star formation, i.e. from the gravitational collapse of a molecular cloud. However, the mechanisms leading to an initial collapse at lower masses (e.g.~turbulent fragmentation;~\citealt{2002ApJ...576..870P}), to formation in different locations (e.g.~disk fragmentation;~\citealt{2009MNRAS.392..413S}), to halt accretion onto a protostar (e.g.~photoevaporation or by ejection of prestellar cores;~\citealt{2004AandA...427..299W,2001AJ....122..432R}, respectively) have not been fully determined. 

The minimum fragmentation mass is determined by the opacity of the gas in a molecular cloud, as high densities cause it to become opaque to its own radiation, and leading to a maximum density at which fragmentation can occur~\citep{1976MNRAS.176..367L,1976MNRAS.176..483R,1977ApJ...214..152S,1977ApJ...214..718S}. From simple critical Jeans mass arguments~\citep{1902RSPTA.199....1J}, \citet{1976MNRAS.176..367L} derived a minimum fragmentation mass of 7\,\mjup, not taking into account effects from magnetic fields, rotation, or late accretion. Magneto-hydrodynamical simulations from~\citet{2012MNRAS.419.3115B} which include these effects, as well as turbulence, shock compression, and radiative feedback, find a lower minimum fragmentation mass of 3\,\mjup~(see also~\citealt{2005AandA...430.1059B,2005MmSAI..76..187P,2007ApJ...661..972P}). Observationally,~\citet{2019ApJS..240...19K} set an initial constraint on the minimum fragmentation mass of 5\,\mjup by fitting a simulated population from evolutionary models to their luminosity functions. However, this result is sensitive to the frequency of coldest Y dwarfs, which only included WISE~J0855$-$0714 at the time of publication. 
WISEA~J0830+2837, CWISEP~J1446$-$2317, and CWISEP~J1935$-$1546 are critical additions to the currently undersampled \teff$\lesssim300$\,K population, and are essential to confidently constrain the low-mass cutoff of the substellar initial mass function. To see whether the theoretically predicted minimum fragmentation masses do indeed describe the observed brown dwarf population, future empirical studies of the initial mass function will need to incorporate the growing sample of ``missing link'' Y dwarfs. However, such calculations are beyond the scope of this work.



\subsection{Characterization of low-temperature atmospheres, from the T/Y dwarf transition to Jupiter}

Our sample contains five ultracool dwarfs, one of which connects the population of known Y dwarfs with the coldest brown dwarf ever discovered. The rest of our targets are most likely late-T dwarfs based on the available evidence. WISEA~J0830+2837 is the $29^{th}$ Y dwarf ever discovered, and the first one from the BYW:P9 collaboration. While most Y dwarfs have been spectroscopically confirmed, their intrinsic faintness in NIR leads to poor S/N and an incomplete characterization of their atmospheric composition.
The best atmospheric analog for these extremely cold brown dwarfs is Jupiter itself. While T dwarfs have atmospheres rich in methane and water, identified by the deep absorption bands in their near-infrared spectra, Y dwarfs are expected to have water vapor clouds~(\citealt{2018ApJ...858...97M,2016ApJ...826L..17S,2014ApJ...793L..16F}). At the T/Y dwarf boundary, Na and K alkalis completely disappear~\citep{2019ApJ...877...24Z}, drastically affecting the $Y-J$ colors at the boundary. At the low temperatures of Y dwarfs, atmospheric models predict the emergence of NH$_3$ in their NIR spectra~\citep{2003ApJ...596..587B,2007ApJ...667..537L,2002Icar..155..393L}, and condensation of salt and sulfide clouds~\citep{2012ApJ...756..172M}. While tentative ammonia absorption has been observed in the NIR spectrum of an object at the T/Y transition~\citep{2012ApJ...745...26B}, and NH$_3$ abundances have been extracted from atmospheric retrievals of Y dwarf NIR spectra as absorption features were attributed to this gas~\citep{2019ApJ...877...24Z}, a conclusive absorption feature matching theoretical predictions is still lacking (e.g., \citealt{2015ApJ...804...92S}). Phosphine, which abounds in the Jovian atmosphere at $4.5-4.6\mu$m, was also expected in Y dwarfs. However, mid-infrared spectra of WISE~J0855$-$0714 did not show indications of this molecule~\citep{2018ApJ...858...97M,2016ApJ...826L..17S}. Therefore, we need higher S/N, higher resolution spectra, and/or a broader wavelength coverage, as well as more developed line lists for these molecules, to identify the components of low-temperature atmospheres. 

Observationally, the spread in colors currently seen at the T/Y dwarf boundary and throughout the Y dwarf class, could be driven by metallicity, gravity, binarity, cloud coverage, or variability effects. A predicted reversal of NIR color trend from bluer in T-dwarfs to redder in Y-dwarfs~\citep{2003ApJ...596..587B,2012ApJ...750...74S,2014ApJ...789L..14M} has been confirmed with a spectral energy distribution of WISE~J0855$-$0714~\citep{2016AJ....152...78L}, and attributed to energy redistribution at longer wavelengths at colder temperatures, and the collapse of the Wien tail~\citep{2016ApJ...823L..35S}.~\citet{2014ApJ...793L..16F} interpreted their J3-[4.5] color measurement as an indication of water ice and sulfide clouds based on equilibrium chemistry models. On the other hand,~\citet{2014ApJ...796....6L} were able to reproduce this color with cloudless, disequilibrium chemistry models, leading to the conclusion that no set of models could simultaneously reproduce the near and mid-infrared photometry~\citep{2016ApJ...823L..35S, 2016AJ....152...78L}. However,~\citet{2016ApJ...826L..17S} and~\citet{2018ApJ...858...97M} disputed the ~\citet{2014ApJ...796....6L} result and validated water ice clouds as the interpretation for the absorption features in the $3.4-5.2\,\mu$m spectra. Finally, photometric variability at a 4\% level was detected in WISE~J0855$-$0714 in the mid-infrared~\citep{2016ApJ...832...58E}, suggesting a patchy atmosphere, yet the cloud  composition remains unclear.

WISEA~J0830+2837, along with CWISEP~J1935$-$1546 and CWISEP~1446$-$2317, fill an important gap in the Y dwarf sequence to continuously map atmospheric composition across low temperatures down to Jupiter. Atmospheric retrievals on a medium resolution, near-infrared spectrum of this object with NIRSpec aboard the \textit{James Webb Space Telescope} would provide the necessary S/N to identify individual absorption lines and bands to characterize their atmospheres.


\section{Conclusions}

We have analyzed \hst and \spitzer photometry of a sample of five ultracool dwarfs identified through the Backyard Worlds: Planet 9 project. We have identified four late-T dwarfs within 30\,pc and one of the coldest Y dwarfs ever recorded, WISEA~J0830+2837. This source has a solid detection in $ch1$ and $ch2$ \spitzer bands, but it drops out on both $F105W$ and $F125W$ \hst bands. These \hst bands are centered in near-infrared wavelengths, where this cold object no longer emits sufficient flux to be detected, while most of its energy is being emitted in longer, mid-infrared wavelengths to which the \spitzer filters are sensitive. 
WISEA~J0830+2837 joins CWISEP~J1935$-$1546 and CWISEP~1446$-$2317 in a small sample of objects serving as a bridge between WISE~J0855$-$0714 and the known population of Y dwarfs. The Backyard Worlds: Planet 9 citizen science project is proving its efficacy at identifying the coldest brown dwarfs, and will continue to provide exceptional targets for follow-up observations.


\acknowledgments

We thank our anonymous referee for their helpful and insightful suggestions that have greatly improved the clarity of this paper. The Backyard Worlds: Planet 9 team would like to thank the many Zooniverse volunteers who have participated in this project, from providing feedback during the beta review stage to classifying flipbooks to contributing to the discussions on TALK. We would also like to thank the Zooniverse web development team for their work creating and maintaining the Zooniverse platform and the Project Builder tools. This research was supported by NASA ADAP grant NNH17AE75I. This publication makes use of data products from the Wide-field Infrared Survey Explorer, which is a joint project of the University of California, Los Angeles, and the Jet Propulsion Laboratory/California Institute of Technology, funded by the National Aeronautics and Space Administration. This research has made use of the NASA/IPAC Infrared Science Archive, which is funded by the National Aeronautics and Space Administration and operated by the California Institute of Technology. This research has made use of the VizieR catalogue access tool, CDS, Strasbourg, France (DOI : 10.26093/cds/vizier). The original description of the VizieR service was published in 2000, A\&A 143, 23. This research made use of APLpy, an open-source plotting package for Python~\citep{2012ascl.soft08017R}.

\vspace{5mm}
\facilities{\hst(WFC3), \spitzer(IRAC), IRSA, AllWISE, CatWISE}

\software{APLpy~\citep{2012ascl.soft08017R}, astrodrizzle~\citep{2012drzp.book.....G},  ~astropy~\citep{2013AandA...558A..33A}, MOPEX/APEX~\citep{2005PASP..117.1113M}, Pandas~\citep{mckinney12}, photutils~\citep{2019zndo...3368647B}, SAOviewerDS9~\citep{2003ASPC..295..489J}, WiseView~\citep{2018ascl.soft06004C}.}

\bibliographystyle{apj}

\end{document}